\documentclass[pra,floatfix,twocolumn,10pt]{revtex4-1} 

\usepackage{graphicx}
\usepackage{amsmath}
\usepackage{amsfonts}
\usepackage{amssymb}
\usepackage{nameref}
\usepackage{amsthm}
\usepackage{hyperref}

\newtheorem{theorem}{Theorem}

\newtheorem{conjecture}{Conjecture}

\newcommand{\eqnref}[1]{Eq.~(\ref{#1})}
\newcommand{\ket}[1]{\ensuremath{|#1\rangle}}
\newcommand{\braket}[2]{\ensuremath{\langle#1|#2\rangle}}

\newcommand{\ketbrad}[1]{\ensuremath{|#1\rangle\langle#1|}}
\newcommand{\matrixel}[3]{\ensuremath{\langle#1|#2|#3\rangle}}
\newcommand{\Tr}{\mathrm{Tr}}
\newcommand{\average}[1]{\big\langle#1\big\rangle}
\newcommand{\onenormc}[1]{\big|#1\big|_1}
\newcommand{\onenormq}[1]{\big|\big|#1\big|\big|_1}

\newcommand{\Uf}{U_f}
\newcommand{\Ufa}{U_{\mathbf{a}}}
\newcommand{\sampf}{\ensuremath{\left|\psi_{f}\right\rangle}}
\newcommand{\sampfa}{\ensuremath{\left|\psi_{\mathbf{a}}\right\rangle}}
\newcommand{\prepstate}{\ket{\Psi_\mathrm{Prep}}}
\newcommand{\preplabel}{\Psi_\mathrm{Prep}}
\newcommand{\hami}{h_f^{(i)}}
\newcommand{\pami}{\pi_f^{(i)}}
\newcommand{\svec}{\mathbf{s}}
\newcommand{\tvec}{\mathbf{t}}
\newcommand{\xvec}{\mathbf{x}}
\newcommand{\vvec}{\mathbf{v}}

\newcommand{\onehalf}{\frac{1}{2}}
\newcommand{\onequarter}{\frac{1}{4}}
\newcommand{\oneninetwo}{\frac{1}{192}}
\newcommand{\oneeightsix}{\frac{1}{86}}
\newcommand{\sqrtonehalf}{\frac{1}{\sqrt{2}}}

\newcommand{\aijk}{a_{ijk}}
\newcommand{\fa}{\mathbf{a}}
\newcommand{\fb}{\mathbf{b}}
\newcommand{\fc}{\mathbf{c}}
\newcommand{\fab}{\mathbf{a}+\mathbf{b}}
\newcommand{\fbc}{\mathbf{b}+\mathbf{c}}
\newcommand{\nb}{n_{\mathbf{b}}}

\newcommand{\ngap}{\mathrm{ngap}}
\newcommand{\ngapsq}{\mathrm{ngap}^2}
\newcommand{\ngapsqest}{\mathrm{ngap}_{\mathrm{Est}}^2}
\newcommand{\qest}{Q_{f,\mathrm{Est}}}
\newcommand{\qcoarse}{\tilde{Q}_{\mathbf{a}+\mathbf{b}}}
\newcommand{\rhocoarse}{\tilde{\rho}_{\mathbf{a}+\mathbf{b}}}
\newcommand{\bigo}{O}
\newcommand{\poly}{\mathrm{poly}}
\DeclareMathOperator{\Prob}{Prob}

\graphicspath{ {./Figures/} }

\begin{document}

\title{Quantum supremacy in constant-time measurement-based computation:\\A unified architecture for sampling and verification}

\author{Jacob Miller}
\email{jmjacobmiller@gmail.com}
\author{Stephen Sanders}
\author{Akimasa Miyake}
\email{amiyake@unm.edu}
\affiliation{Center for Quantum Information and Control, Department of Physics and Astronomy, University of New Mexico, Albuquerque, NM 87131, USA}

\begin{abstract}

While quantum speed-up in solving certain decision problems by a fault-tolerant universal quantum computer has been promised, a timely research interest includes how far one can reduce the resource requirement to demonstrate a provable advantage in quantum devices without demanding quantum error correction, which is crucial for prolonging the coherence time of qubits. We propose a model device made of locally-interacting multiple qubits, designed such that simultaneous single-qubit measurements on it can output probability distributions whose average-case sampling is classically intractable, under similar assumptions as the sampling of non-interacting bosons and instantaneous quantum circuits. Notably, in contrast to these previous unitary-based realizations, our measurement-based implementation has two distinctive features.  (i) Our implementation involves no adaptation of measurement bases, leading output probability distributions to be generated in constant time, independent of the system size. Thus, it could be implemented in principle without quantum error correction. (ii) Verifying the classical intractability of our sampling is done by changing the Pauli measurement bases only at certain output qubits. Our usage of random commuting quantum circuits in place of computationally universal circuits allows a unique unification of sampling and verification, so that they require the same physical resource requirements in contrast to the more demanding verification protocols seen elsewhere in the literature. 

\end{abstract}

\maketitle

\section{Introduction}

General-purpose quantum computers hold the promise of achieving quantum speed-ups in many problems of practical importance, unmatched by any known classical methods \cite{shor1997polynomial, grover1997quantum, deutsch1992rapid}. While the prospect of such speed-ups is exciting, a growing realization is the extreme difficulty of achieving the levels of precision and control required for building truly scalable, fault-tolerant quantum hardware. As an intermediate step towards this goal, several recent proposals have suggested the development of special-purpose quantum devices which achieve so-called ``quantum supremacy" in certain tasks \cite{terhal2004adaptive, bremner2011classical, aaronson2013computational, morimae2014hardness, bremner2016average, rahimikeshari2016sufficient, farhi2016quantum, rohde2016quantum, boixo2016characterizing, gao2017quantum, fujii2016noise, bremner2016achieving, lund2017sampling, bermejovega2017architectures, fefferman2017exact, shahandeh2017quantum, deshpande2017complexity, kapourniotis2017nonadaptive}. Instead of solving general computational problems, these devices instead sample from probability distributions widely believed to be impossible to simulate efficiently using classical means. The recent explosion of proposals for such classically intractable sampling devices has begun to be matched by actual demonstrations of sampling in the laboratory \cite{spring2013boson, tillmann2013experimental, crespi2013integrated, broome2013photonic, wang2016multi}, although so far still at small enough scales to allow for exact classical simulation.

An important question regarding such proposals is how far, and in what manner, we can reduce the resources required to exhibit and certify a genuine quantum advantage in sampling. The boson sampling protocol \cite{aaronson2013computational} shows that such quantum advantage can be achieved using simple linear optical devices and single-photon detectors. However, there are many challenges facing a realistic implementation of boson sampling, including the parallel generation of many single photons, the precise timing constraints on these photons, and the robust and accurate arrangement of the required beam splitters and phase shifters. An alternative proposal which circumvents this bottleneck is the family of instantaneous quantum polynomial-time (IQP) protocols \cite{bremner2011classical, bremner2016average, bremner2016achieving}, where sampling distributions arise from single-qubit measurements on the output of low-depth commuting quantum circuits. If a quantum device can prepare sampling distributions associated with any unitary within a circuit family, then that process would be classically intractable under reasonable conjectures from complexity theory. Furthermore, the commuting nature of these quantum circuits means that they can potentially be engineered to run in constant time, maximally avoiding the threat of environmental noise and decoherence. However, a practical issue which arises here is the extreme difficulty of engineering the arbitrary long-range interactions needed for such a constant time implementation. While these long-range interactions can be simulated by bringing distant qubits together using $SWAP$ gates before applying local entangling operations, this process would introduce a new bottleneck, the growing time required to shuttle qubits between local interaction regions. In the absence of quantum error correction, the growing influence of decoherence would quickly degrade the quality of our sampling distributions, making this straightforward implementation likely untenable for practical demonstrations of quantum supremacy.

In this paper, we show how nonadaptive measurement-based quantum computation (MQC) \cite{raussendorf2001one, jozsa2005introduction, briegel2009measurement} can be used to sample from the distributions associated with IQP circuits, while at the same time verifying the classical intractability of this sampling process. Our protocol uses a fixed resource state preparable by a constant-depth local circuit, which is then nonadaptively measured at each site in the Pauli $X$, $Y$, or $Z$ bases. The setting of nonadaptive MQC allows us to replace the time complexity present in local IQP circuits (with $SWAP$ gates) by a spatial overhead in our resource state, which results in a protocol with constant runtime and local interactions. The cost of this nonadaptivity is a fundamental randomness in the distributions prepared by our protocol, arising from random MQC byproduct operators. This leads each sample in our protocol to be obtained with high probability from a different sampling distribution every time. Surprisingly, we show that this inherent randomness has no impact on the hardness of our protocol, which remains classically intractable under the same assumptions as in \cite{bremner2016average}. What's more, we show that these random byproduct operators actually simplify our implementation relative to a direct circuit-based counterpart, revealing an inherent advantage of MQC for quantum sampling protocols. We further show that by simply changing the single-qubit Pauli measurements used in the final step of our protocol to obtain sampling statistics, we can instead rigorously verify the classical intractability of our sampling. Our verification scheme is inspired by the ground state certification protocol of \cite{hangleiter2017direct}, but uses the special form of our IQP sampling distributions to replace the nonlocal operations required for general Hamiltonian measurements with measurements of single-qubit Pauli operators. This lets us switch between sampling and measurement by a simple change in single-qubit measurement bases, allowing our procedure to achieve a robust demonstration of quantum supremacy capable of efficiently detecting any errors which could potentially harm the correctness of our sampling distributions.

Our protocol is closely related to that of \cite{bremner2016average}, as it constitutes a faithful translation of their circuit-based IQP sampling into the context of MQC. However, we show that this translation itself contains several surprises, ultimately revolving around the nontrivial interface of MQC byproduct operators with classically intractable sampling. At first glance, our protocol has much in common with \cite{gao2017quantum, bermejovega2017architectures}, which also use nonadaptive MQC to perform classically intractable sampling and verification. Upon further investigation however, the different protocols are seen to utilize completely different mechanisms for demonstrating quantum supremacy, which allow for substantial differences in behavior. While using a more involved resource state than the Ising-like states of \cite{gao2017quantum, bermejovega2017architectures}, the design of our protocol allows a unique duality between sampling and verification, in that both require the same physical resources and are switchable by a mere change of single-qubit Pauli measurement bases on an $n$-qubit output state. This feature fundamentally depends upon the convenient mathematical nature of our IQP sampling distributions, and cannot be straightforwardly reproduced within the setting of sampling from random universal circuits such as \cite{gao2017quantum, bermejovega2017architectures}.

In Section~\ref{sec:background}, we review the relevant theory behind IQP sampling, verification, and MQC. In Section~\ref{sec:protocol} we present our protocol for preparing, sampling from, and verifying different classically intractable sampling distributions using Pauli measurements on a model resource state $\prepstate$. In Section~\ref{sec:outlook} we comment on the features unique to our protocol, and outline future directions for our work. A brief comparison of our proposal to other proposals within the rapidly growing field of classically intractable sampling can be found in Appendix~\ref{sec:comparison}, with detailed proofs of the classical intractability and verification of our sampling protocol found in Appendices~\ref{sec:preparation}, \ref{sec:sampling} and \ref{sec:verification}.

\section{Background}
\label{sec:background}
\subsection{IQP and Boolean Functions}
\label{sec:iqp}

In the IQP sampling protocols of \cite{bremner2011classical, bremner2016average, bremner2016achieving}, a sampling state $\sampf = \Uf \ket{+}^{\otimes n}$ is first prepared using an $n$-qubit diagonal unitary circuit $\Uf$, and is then measured everywhere in the Pauli $X$ basis to obtain a random outcome $\ket{\svec_X}=H^{\otimes n} \ket{\svec}$. In the above, $\ket{+}=\sqrtonehalf(\ket{0}+\ket{1})$ denotes the $+1$ eigenstate of $X$, $H$ the single-qubit Hadamard operator, $\svec=(s_1,s_2,\ldots,s_n)$ a bit string of length $n$, and $\ket{\svec}$ the corresponding $Z$ basis product state. If $\Uf$ is chosen from an appropriate family of diagonal unitaries, then \cite{bremner2011classical} shows that the act of sampling from the distribution $D_{f}(\svec) = |\braket{\svec_X}{\psi_f}|^2$ is impossible to perform in polynomial time using a classical computer, assuming the widely conjectured non-collapse of the polynomial hierarchy of complexity theory \cite{meyer1972equivalence, stockmeyer1977polynomial}. More generally, we use the phrase classically intractable sampling to mean any sampling protocol which shares this property of being impossible to simulate classically (given the non-collapse of the polynomial hierarchy), possibly in the presence of some allowable error and under the assumed truth of additional mathematical conjectures.

We now choose the $n$-qubit unitary gates $\Uf$ above to be parameterized by $n$-bit binary functions $f:GF(2)^n\to GF(2)$, where $GF(2)\simeq\{0,1\}$ denotes the finite field of binary numbers. The functions $f$ set the eigenvalues of $\Uf$ as
\begin{equation}
    \Uf = \sum_{\xvec\in GF(2)^n} (-1)^{f(\xvec)} \ketbrad{\xvec},
\end{equation}
where $\xvec=(x_1,x_2,\ldots,x_n)$. When applied to $\ket{+}^{\otimes n}$, this results in the sampling state
\begin{equation}
    \sampf = 2^{-n/2} \sum_{\xvec\in GF(2)^n} (-1)^{f(\xvec)} \ket{\xvec}.
\end{equation}
We can alternately describe $\sampf$ as the unique state satisfying the $n$ (nonlocal) stabilizer relations $\hami\sampf = (+1)\sampf$ for $1\leq i\leq n$, where
\begin{align}
\label{eq:hami}
    \hami &= \Uf X_i\Uf^\dagger \nonumber\\
    &= X_i \sum_{\xvec\in GF(2)^n} (-1)^{\partial_{i}f(\xvec)} \ketbrad{\xvec},
\end{align}
and the polynomial $\partial_{i}f$ is equal to the difference
\begin{equation}
\label{eq:partial}
    \partial_{i}f(\xvec) = f(x_1,\ldots,x_i+1,\ldots,x_n)-f(x_1,\ldots,x_i,\ldots,x_n).
\end{equation}
Because addition in $GF(2)$ is modulo 2, it is easy to verify that $\partial_{i}f(\xvec)$ is always independent of the value of $x_i$.

We now restrict our binary functions to be cubic polynomials, so that $f(\xvec)$ can be written in the form
\begin{multline}
    \label{eq:cubic_poly}
    \!\!\!\!\!f(\xvec) = \!\!\!\!\!\sum_{1\leq i<j<k\leq n} \!\!\!\!\aijk x_i x_j x_k\,\,\, + \!\!\sum_{1\leq i<j\leq n} \!\!b_{ij} x_i x_j\,\, + \!\sum_{1\leq i\leq n} \!c_{i} x_i,
\end{multline}
for some binary coefficients $\aijk$, $b_{ij}$, and $c_{i}$. These are generated by linear, quadratic, and cubic monomials, whose associated diagonal unitary gates are $U_{x_i}=Z_i$, $U_{x_i x_j}=CZ_{ij}$ (controlled-$Z$), and $U_{x_i x_j x_k}=CCZ_{ijk}$ (controlled-controlled-$Z$). More explicitly, the gates $Z_i$, $CZ_{ij}$, and $CCZ_{ijk}$ are defined by their action on qubits $i$, $j$, and $k$ respectively as $Z_i \ket{x_i} = (-1)^{x_i} \ket{x_i}$, $CZ_{ij} \ket{x_i,x_j} = (-1)^{x_ix_j} \ket{x_i,x_j}$, and $CCZ_{ijk} \ket{x_i,x_j,x_k} = (-1)^{x_i x_j x_k} \ket{x_i,x_j,x_k}$. In the following, any references to polynomials will be understood to refer specifically to binary polynomials. We will use $\fa$, $\fb$, and $\fc$ to denote homogeneous polynomials, for which the only nonzero coefficients are of the form $\aijk$, $b_{ij}$, or $c_{i}$, respectively. Similarly, $\fbc$ and $\fab$ will denote polynomials for which all $\aijk=0$ or all $c_i=0$, respectively.

It will be convenient in the following to interpret $n$-bit vectors $\svec$ as linear polynomials of $n$ variables, which act as
\begin{equation}
    \svec(\xvec) = \sum_{i=1}^n s_i x_i.
\end{equation}
This is useful in giving the probability of different sampling outcomes, as the probability of obtaining any given $\ket{\svec_X}$ when $\sampf$ is measured in the $X$ product basis is
\begin{align}
\label{eq:distribution}
    D_{f}(\svec) &= \big|\braket{\svec_X}{\psi_f}\big|^2 \nonumber\\
    &= \bigg|2^{-n} \sum_{\xvec\in GF(2)^n} (-1)^{f(\xvec)+\svec(\xvec)} \bigg|^2 \nonumber\\
    &= \ngapsq(f+\svec).
\end{align}
$\ngapsq(f)$ refers here to the square of $\ngap(f)$, the (signed) difference between the fraction of inputs yielding $f(\xvec)=0$ and $f(\xvec)=1$. $\ngap(f)$ is known to be \#P-hard to compute for arbitrary cubic polynomials $f$ \cite{ehrenfeucht1990computational}, and we will see that this hardness underlies the classical intractability of our sampling protocol.

\subsection{Classically Intractable Sampling and Verification}
\label{sec:intractable}

It is shown in \cite{bremner2016average} that estimating the quantity $\ngapsq(f)$ up to $\onequarter$ multiplicative error, so that $|\ngapsqest(f)-\ngapsq(f)|\leq \onequarter\ngapsq(f)$ for arbitrary cubic polynomials $f$, is \#P-hard, mirroring the difficulty of computing $\ngap(f)$. This hardness leads to a similar finding as in \cite{bremner2011classical}, that exactly sampling from the cubic polynomial distributions $D_{f}$ defined in \eqnref{eq:distribution} is classically intractable. In particular, assuming the existence of a classical randomized algorithm which can efficiently sample from any of the distributions $D_{f}$ lets a technique called Stockmeyer approximate counting \cite{stockmeyer1985approximation} be used to estimate the probabilities $D_f(\svec)$ up to $\onequarter$ multiplicative error, and thus to solve arbitrary \#P problems. While Stockmeyer counting is an unphysical process which cannot be implemented with realistic classical or quantum computers, it can be carried out at a finite level of the polynomial hierarchy. The hardness of \#P problems for the polynomial hierarchy then leads to its collapse. Details of this process can be found in Appendix~\ref{sec:sampling}. On the other hand, we have seen that these distributions appear naturally as the output distributions of the IQP sampling protocol described above, which allows us to interpret a concrete implementation of this protocol as a provable demonstration of ``quantum supremacy".

While straightforward and conceptually compelling, a major limitation of the above result is the impossibility of verifying that any realistic quantum protocol is sampling from \textit{exactly} the ideal distribution $D_{f}$ \cite{footnote1}. In order to demonstrate quantum supremacy in a more realistic setting, an alternate proof is given in \cite{bremner2016average} which shows the classical intractability of sampling from any distribution $Q_f$ which is variationally close to $D_f$. Variationally close means here that the statistical distance between $Q_f$ and $D_f$ is bounded by a constant $\eta_0$, so that
\begin{equation}
    \label{eq:epsilon}
    \onenormc{Q_f-D_f} = \!\!\!\sum_{\svec\in GF(2)^n} |Q_f(\svec)-D_f(\svec)|\leq\eta_0,
\end{equation}
In \cite{bremner2016average} a value of $\eta_0\leq\oneninetwo$ was shown to be sufficient for classically intractable sampling, which in Appendix~\ref{sec:sampling} we show can be relaxed to $\eta_0\leq\oneeightsix$ (although both values rely on the particular value of $\epsilon_0$ appearing in Conjecture~\ref{conj} below). This result is appealing from a practical standpoint, as the quantity $\onenormc{Q_f-D_f}$ can be efficiently estimated in experiments involving quantum sampling distributions.

On the other hand, the above ``average-case" sampling result relies upon one additional complexity theoretic conjecture:
\begin{conjecture}[Average-Case Hardness of $\ngapsq(f)$]
    \label{conj}
    Let $f$ be an arbitrary cubic polynomial of the form given in \eqnref{eq:cubic_poly}. Then it is \#P-hard to efficiently calculate an estimate $\ngapsqest(f)$ of $\ngapsq(f)$ for which $|\ngapsqest(f)-\ngapsq(f)|\leq\onequarter\ngapsq(f)$, on at least $1-\epsilon_0=\frac{1}{24}$ of polynomials $f$.
\end{conjecture}
\noindent Intuitively, this conjecture states that even when our estimates $\ngapsqest(f)$ are allowed to fail with some finite probability $\epsilon_0$, corresponding to realistic errors in our sampling distributions $Q_f$, the problem of estimating $\ngapsq(f)$ on the remaining instances is still \#P-hard. While this reliance on an additional unproven conjecture isn't desirable, an analogous conjecture is required for every known average-case classically intractable sampling result, and thus isn't any special demerit of \cite{bremner2016average}.

The techniques of \cite{hangleiter2017direct} can be used to efficiently verify the condition $\onenormc{Q_f-D_f}\leq\eta_0$ when $Q_f$ arises from measurements on experimentally prepared quantum sampling states $\rho_f$, which approximate our intended $\sampf$. Given $\rho_f$, we can perform measurements of the nonlocal Hermitian stabilizers $\hami$ defined in \eqnref{eq:hami}, which will always yield the outcome $+1$ in the ideal case where $\rho_f=\ketbrad{\psi_f}$. In more general cases, a sufficiently accurate empirical estimate of these $n$ observables $\hami$ can be converted into a bound on the statistical distance between the distributions $Q_f$ and $D_f$. If the average $\average{\hami}$ is sufficiently close to $+1$ so as to guarantee $\onenormc{Q_f-D_f}\leq\eta_0$, then we can confidently conclude that our quantum protocol is performing classically intractable sampling. We will soon show that the nonlocal measurements of $\hami$ can actually be entirely replaced with single-qubit $X$ and $Z$ measurements, which allows this verification to be done within the setting of MQC.

\subsection{Measurement-Based Quantum Computation}
\label{sec:mqc}

\begin{figure}[t]
    \centering
    \includegraphics[width=0.49\textwidth]{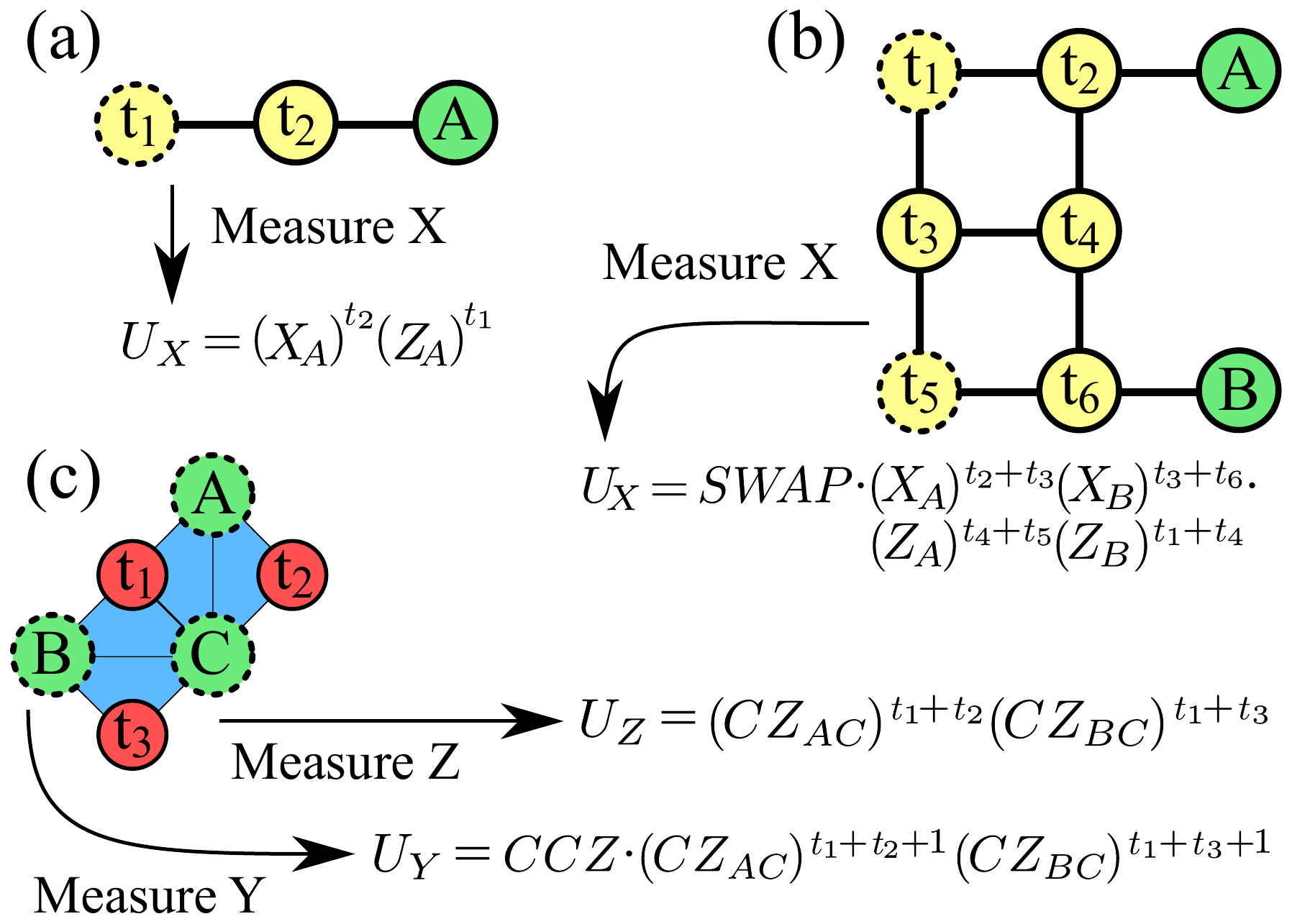}
    \caption{The MQC gadgets utilized in our protocol. We describe the formation circuit and outcome-dependent logical operations for each when sites are measured everywhere in some single-qubit Pauli basis. Initial states on input sites (dotted) are teleported to output sites (green) and acted on by characteristic logical operations and measurement-dependent random byproduct operators. These outputs are identified with the inputs of future gadgets. The relationship between the formation circuits shown and the logical operations implemented is given by contracting with the appropriate measurement outcomes, which additionally contributes a scalar factor of $\sqrtonehalf$ per measurement (not shown). Mathematically, this leads our measurement outcomes $s_i$ to occur uniformly randomly. (a) 1D cluster state wire of length 2, where solid lines indicate $CZ$ formation unitaries. Measuring $X$ on two sites implements the identity, with a uniformly random Pauli byproduct group. (b) Planar MQC gadget for implementing nonplanar wire crossings. Measuring $X$ on 6 sites implements $SWAP$, with a byproduct group of uniformly random two-qubit Pauli operators. (c) Non-Clifford gadget for conditional $CCZ$, where blue triangles indicate $CCZ$ gates used to form the gadget. Measuring $Y$ on 3 non-logical control sites (dark red, smaller circles) gives $CCZ$ on sites $A$, $B$, and $C$, whereas measuring $Z$ on these sites instead gives the identity. In both cases, the teleportation is trivial (output and input sites coincide), while the byproduct group is a product of uniformly random $CZ$'s between $A$ and $C$, and between $B$ and $C$.}
    \label{fig:gadgets}
\end{figure}

MQC is a means of carrying out computation using only single-qubit measurements on a fixed many-body resource state. In this framework, the choice of measurements made on local regions of our resource state determines logical operations which are applied to encoded logical qubits, while simultaneously teleporting these qubits to adjacent unmeasured sites. The randomness of quantum measurement leads the outcomes of these measurements to determine a so-called byproduct operator, which acts as a random correction to the overall logical operation. For example, in Figure~\ref{fig:gadgets}a we show the standard protocol for teleporting one logical qubit within the MQC quantum wire known as the 1D cluster state. Given two successive $X$ measurements with outcomes $\ket{t_{1,X}}$ and $\ket{t_{2,X}}$, the resultant logical operation is $U_X(t_1,t_2) = X^{t_2}Z^{t_1}$, showing the intended logical unitary to be the identity and the byproduct operator to be a random Pauli $X^{t_2}Z^{t_1}$. In Figure~\ref{fig:gadgets}b we show a gadget for performing the two-qubit $SWAP$ operation on logical qubits, for which the byproduct operator is a random two-qubit Pauli operator. In both of these examples, the collection of operators appearing as byproducts for arbitrary measurement outcomes form a closed group (up to global phase) of finite size, referred to as a byproduct group.

An MQC protocol is said to be adaptive if the choice of measurement in some region of our resource state depends on the outcome of measurements made in another region. Adaptation can be seen as a means of ensuring that the byproduct group associated with a large computation remains finite (for example, contained within the $n$-qubit Pauli group), whereas the use of nonadaptive MQC with arbitrary single-qubit measurements will generally lead to a byproduct group of unbounded size. On the other hand, nonadaptive MQC computations can always be implemented in constant time by performing all measurements simultaneously, a serious advantage in the absence of quantum error correction. Within the usual scheme for universal MQC using resource states built from $CZ$ gates, nonadaptive single-qubit Pauli measurements are associated with byproduct groups formed from Pauli operators, and implement logical operations contained within the Clifford group. The Clifford group is defined as those unitaries $U$ which preserve the Pauli group under conjugation, so that $UPU^\dagger$ is a product of Pauli operators whenever $P$ is. The evolution of Pauli eigenstates under the Clifford group is known to be efficiently simulable using classical means \cite{gottesman1998heisenberg}, which means that non-Clifford operations are necessary for demonstrating quantum supremacy.

In Figure~\ref{fig:gadgets}c, we give an example of an MQC gadget which implements a non-Clifford $CCZ$ gate when nonadaptive Pauli measurements are applied. This gadget, which will be utilized in our classically intractable sampling protocol below, is itself formed from non-Clifford $CCZ$ gates, and has a byproduct group containing non-Pauli $CZ$ gates. A similar gadget was shown in \cite{miller2016hierarchy} to enable universal MQC using only Pauli measurements, but with adaptation of measurement bases so as to avoid a byproduct group of unbounded size. In our MQC sampling protocol below, we will show that restricting our logical operations to those generating sub-universal quantum computation will allow us to avoid this use of adapation, while still maintaining a byproduct group of finite size. In fact, we will find that this non-Pauli byproduct group actually leads to a simplification in our protocol relative to circuit-based counterparts.

\section{MQC Protocol for Classically Intractable Sampling}
\label{sec:protocol}

\begin{figure*}[t]
    \centering
    \includegraphics[width=0.9\textwidth]{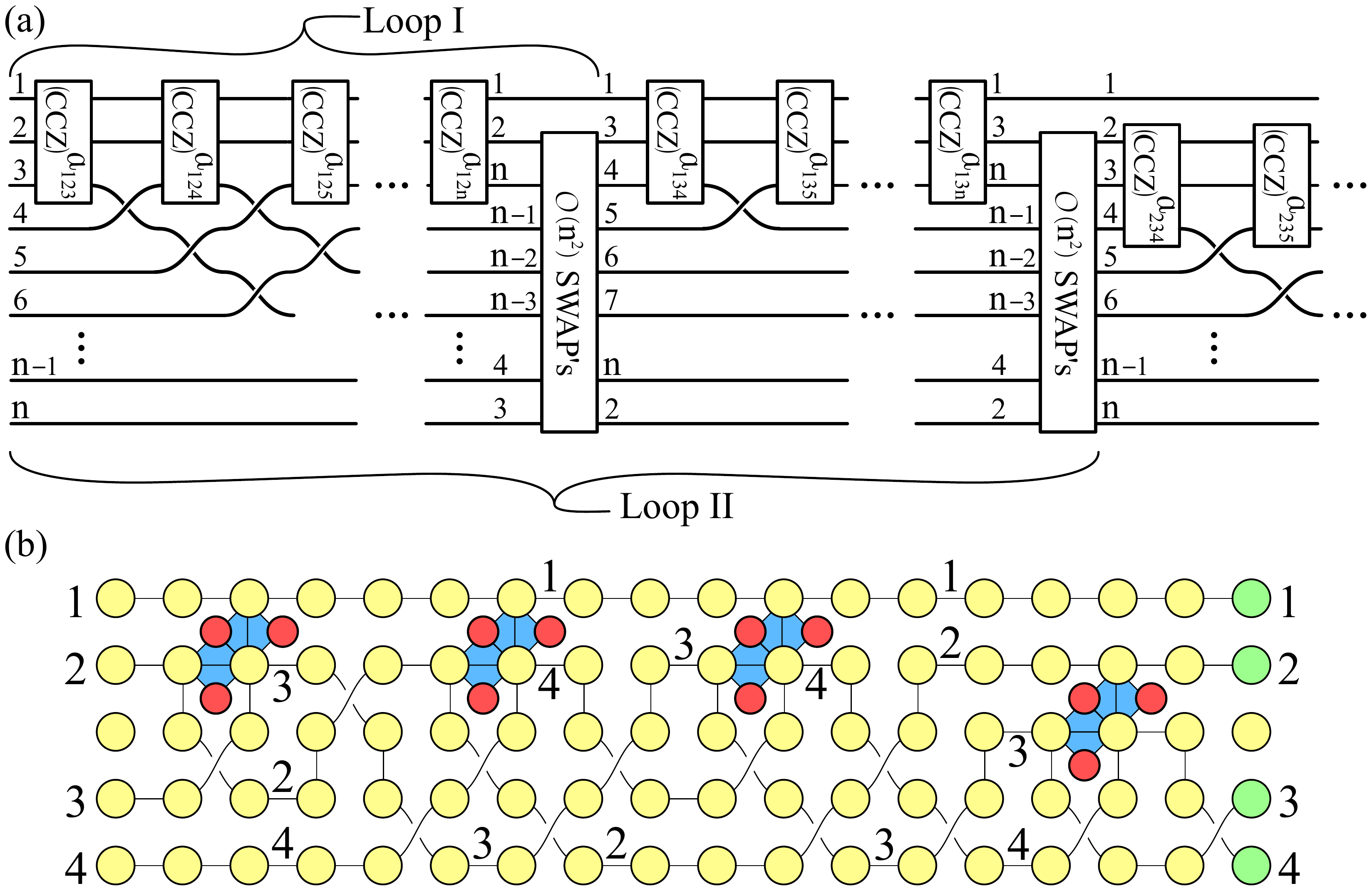}
    \caption{An overview of our constant-time MQC protocol for implementing the unitary $\Uf=U_{\fbc}\Ufa$ which prepares the sampling state $\sampf$. Our intended logical operation is $\Ufa$, while $U_{\fbc}$ is a byproduct contribution containing uniformly random $\fb$ and $\fc$. (a) Circuit diagram for $\Ufa$, which is formed from several repeating loops. In Loop I, qubits $i_0$ and $j_0$ remain fixed and all qubits $k>i_0,j_0$ are sequentially cycled past $i_0$ and $j_0$ and acted on by a conditional three-body gate $(CCZ_{i_0j_0k})^{a_{i_0j_0k}}$ depending on the binary coefficient $a_{i_0j_0k}$ in $f$. The order of these qubits is reversed after Loop I, which is undone by a sequence of $SWAP$'s with circuit depth $\bigo(n)$. Loop II then involves replacing qubit $j_0$ by $j_0+1$, and repeating Loop I for all triples $(i_0,j_0+1,k)$, where $k>i_0,j_0+1$. Loop II continues cycling qubit $j$ and applying Loop I until all triples $(i_0,j,k)$ have been addressed. Loop III (not shown) then involves replacing qubit $i_0$ by $i_0+1$, and repeating Loop II for all triples $(i_0+1,j,k)$. At the completion of Loop III, we have addressed all triples of qubits within circuit depth $\bigo(n^3)$, producing the output state $\ket{\psi_{\fa}}$. (b) A concrete example of how the above protocol is implemented in MQC using our resource state $\prepstate$, for $n=4$. 1D cluster state wires let us teleport information between non-Clifford gadgets, which apply the logical gate $(CCZ)^{\aijk}$ via an $\aijk$-dependent choice of $Y$ or $Z$ measurement on control sites (dark red, smaller circles). While our state is drawn with nonplanar wire crossings, these are simulated using the planar $SWAP$ gadgets in Figure~\ref{fig:gadgets}b. Measuring all preparation sites simultaneously prepares a random $n$-qubit state $\sampf$ on the output sites (on right, green), where $f=\fa+\fbc$ contains a deterministic $\fa$ set by the measurement bases and a uniformly random $\fbc$ arising from random byproduct operators. The final $n$-qubit measurement is chosen to randomly implement sampling via all $X$ measurements, or verification via a mixture of $X$ and $Z$ measurements.}
    \label{fig:protocol}
\end{figure*}

Our MQC implementation of the classically intractable sampling protocol of \cite{bremner2016average} uses nonadaptive Pauli measurements to prepare, sample from, and verify the $n$-qubit sampling states $\sampf$ described above, for arbitrary cubic polynomials $f$. Our protocol uses a 2D resource state $\prepstate$ which is capable of preparing any sampling state $\sampf$ using only single-qubit Pauli measurements. $\prepstate$ is constructed from the teleportation, $SWAP$, and $CCZ$ gadgets described in Section~\ref{sec:mqc}, which are configured to implement any of the IQP circuits $\Ufa$ associated with arbitrary homogeneous cubic polynomials $\fa$. The choice of $\fa$ is determined by the choice of Pauli measurement basis applied to each $CCZ$ gadget in $\prepstate$. By virtue of the byproducts arising from our nonadaptive MQC implementation, our output sampling states end up being random $\sampf$ where $f=\fa+\fbc$ is a sum of the intended $\fa$, along with random quadratic and linear polynomials $\fb$ and $\fc$. Owing to this randomness in $\fbc$, we are unable to deterministically prepare any fixed sampling state $\sampf$. Despite this fundamental indeterminism, we will show how the act of sampling from randomly prepared $\sampf$ with $X$ measurements at the final stage of our protocol remains classically intractable, even in the presence of realistic noise which leads our output sampling distributions to be some imperfect $Q_f$. We state the classical intractability of our protocol, and the precise conditions which guarantee this, as Theorem~\ref{thm:one}.

\begin{theorem}
    \label{thm:one}
    Assume the validity of Conjecture~\ref{conj} and the non-collapse of the polynomial hierarchy. If the distributions $Q_f(\svec)$ arising from our MQC sampling protocol are close on average to the distributions $D_f(\svec)$ defined in \eqnref{eq:distribution}, meaning that the average $\ell_1$ norm over all $f$ meets the experimental threshold $\average{\onenormc{Q_f-D_f}}_f \leq \eta_0=\oneeightsix$, then our protocol is impossible to efficiently simulate using a classical computer, i.e. is classically intractable.
\end{theorem}

Our protocol for classically intractable sampling is divided into two stages: preparation of the random sampling state $\sampf$, followed by sampling and verification measurements on $\sampf$ (see Figure~\ref{fig:protocol}). In the preparation stage, we use $m=\bigo(n^4)$ single-qubit measurements of Pauli $X$, $Y$, and $Z$ on $\prepstate$ with outcomes $\tvec=(t_1,t_2,\ldots,t_m)$ to prepare the $n$-qubit state $\ket{\psi_{f(\tvec)}}$ associated with a $\tvec$-dependent polynomial $f(\tvec)=\fa+\fb(\tvec)+\fc(\tvec)$. These measurements are chosen to implement the unitary $\Ufa$ by means of a depth $\bigo(n^3)$ quantum circuit built from local $CCZ$ and $SWAP$ gates. The $CCZ$ gates in this ideal circuit are applied conditionally as $(CCZ)^{\aijk}$, depending on the coefficients of $\fa$, with teleportation and $SWAP$ gates used before each application to move qubits $i$, $j$, and $k$ into the same region. The application of these conditional $CCZ$'s is structured within three nested levels of iteration, which together apply all $n\choose3$ three-body terms in the lexicographic order of the triples $(i,j,k)$, where $i<j<k$. Loop I, the lowest level of iteration, involves fixing qubits $i$ and $j$ in a designated interaction region, then successively cycling the remaining qubits $k>j$ through this region. $(CCZ)^{\aijk}$ is applied in turn to each triple, until all triples $(i,j,k)$ with fixed $i$ and $j$ have been processed in this manner. Loop II, the next level of iteration, involves successively replacing qubit $j$ by qubit $j+1$, then repeating Loop I for all qubits $k>j+1$ until all triples $(i,j,k)$ with fixed $i$ have been processed. Finally, Loop III involves successively replacing qubit $i$ by qubit $i+1$, in the process shifting the location of the interaction region, and repeating Loop II for all qubits $j,k>i+1$ until $(CCZ)^{\aijk}$ has been applied to all triples of qubits. The resulting unitary operation is clearly $\Ufa$.

While the simple circuit described above is only capable of producing sampling states $\sampfa$ associated with homogeneous cubic $\fa$, our MQC implementation utilizes random byproduct operators to implement the remaining quadratic and linear terms required for the preparation of arbitrary $\sampf$. This reveals a simplification within nonadaptive MQC compared to a direct circuit-based counterpart, which would require additional $CZ$ and $Z$ gates to implement $\Uf$ for arbitrary $f$. Each of the conditional operations $(CCZ)^{\aijk}$ is implemented using the $CCZ$ gadget shown in Figure~\ref{fig:gadgets}c, which is measured in $Y$ if $\aijk=1$ and $Z$ otherwise. For either choice of measurement, the non-Clifford nature of these gadgets leads the resultant byproduct operators to consist of non-Pauli $CZ$ gates, which generate random quadratic terms in the output $\sampf$. Because our logic gates and byproduct operators are made up of $X$ and the diagonal $Z$, $CZ$, and $CCZ$ gates, which together form a closed (non-universal) gate set under multiplication, the byproduct group associated with our computation will always remain finite. This is in contrast to the byproduct group appearing in MQC implementations of random circuit quantum supremacy protocols, such as \cite{gao2017quantum, bermejovega2017architectures}, which grows unboundedly.

The $CCZ$ gadgets used in our protocol are embedded in regular intervals in $\prepstate$, and are then connected together using 1D cluster wires and $SWAP$ gadgets, which simulate the movement of qubits utilized in our ideal quantum circuit described above. These cluster wires and $SWAP$ gadgets are always measured in $X$, which leads to a product of random Pauli $X$ and $Z$ byproduct operators. The $Z$ byproducts eventually end up generating random linear terms in the output state $\sampf$, while the $X$ byproducts can be commuted backwards in our circuit, to eventually be annihilated on the initial $\ket{+}^{\otimes n}$ which our logical quantum circuit is applied to. This commutation of $X$ byproduct operators induces conditional $(CZ)^{\aijk}$ and $(Z)^{\aijk}$ byproduct operators arising from prior $CCZ$ gadgets, which results in additional randomness in the overall byproduct group. Despite this seeming complexity in the distribution of byproduct operators, we prove in Appendix~\ref{sec:preparation} that the random outcomes $\tvec$ of preparation measurements on $\prepstate$ lead the random quadratic and linear terms in the polynomial $f=\fa+\fbc$ associated with $\sampf$ to be uniformly random, simplifying our analysis.

In the second stage of our protocol we apply a final series of $n$ single-qubit Pauli measurements to our output state which, while ideally equal to $\ketbrad{\psi_f}$, will realistically be some mixed state $\rho_f$. The choice of single-qubit measurement bases depends on whether we are implementing sampling or verification, which can be chosen randomly with $\onehalf$ probability. During sampling, we simply measure all qubits in the $X$ basis to generate a sample from the distribution $Q_f(\svec)=\Tr(\rho_f\ketbrad{\svec_X})$, exactly as described in Section~\ref{sec:intractable}. Although the randomness in the $f$ associated with $\rho_f$ means that we will almost certainly obtain each sample from a different distribution $Q_f$, our MQC sampling protocol remains classically intractable nonetheless. To prove this classical intractability, we can treat the overall process of preparing a random $\rho_f$ and then sampling an outcome $s$ as itself a sampling process with probability $\Prob_{\fa}(\fbc,\svec)$. Given this description, and our knowledge of the complete randomness of the byproduct contributions $\fbc$, Stockmeyer approximate counting can then be used to estimate $Q_f(\svec)$ as a conditional probability which is directly proportional to $\Prob_{\fa}(\fbc,\svec)$. This suffices to proves Theorem~\ref{thm:one} using the same arguments as in other classically intractable sampling proposals, the details of which are given in Appendix~\ref{sec:sampling}.

If we choose to perform verification instead of sampling, then we measure all qubits in the $Z$ basis, except for a random qubit $i$ which is measured in $X$. The outcome of this measurement $\vvec=(v_1,v_2,\ldots,v_n)$ is then fed into a parity function $\pami(\vvec) = \partial_{i}f(\vvec)+v_i$, where $\partial_{i}f(\vvec)$ is defined in \eqnref{eq:partial}. This process results in an output value of 0 or 1, which we show in Appendix~\ref{sec:verification} gives the same information as a measurement of the nonlocal stabilizer $\hami$ described in \eqnref{eq:hami}, with outcome $(-1)^{\pami(\vvec)}$. Because of our ability to characterize the closeness of $\rho_f$ to $\ketbrad{\psi_f}$ using measurements of $\hami$, this means that we can interpret $\pami(\vvec)=0$ as a successful verification measurement, and $\pami(\vvec)=1$ as a deviation of $\rho_f$ from our intended $\sampf$. By obtaining many samples of $\pami(\vvec)$ for random $i$, $\vvec$, and $\rho_f$, the resultant estimate of $\average{\pami}$ lets us guarantee the classical intractability of our MQC sampling protocol to any desired statistical significance using only $\bigo(n^2)$ rounds of verification measurements, as stated in Theorem~\ref{thm:two}.

\begin{theorem}
    \label{thm:two}
    Suppose that the empirical average of our parity function after $\mu n^2$ verification measurements satisfies $\average{\pami(\vvec)}_{\vvec,i,f}\leq \frac{\eta_0^2}{n}$, for the $\eta_0$ appearing in Theorem~\ref{thm:one}. Then we can conclude with probability $p\geq1-e^{-\bigo(\mu^2)}$ that our sampling distributions $Q_f$ satisfy the assumptions of Theorem~\ref{thm:one}, and thus generate classically intractable sampling.
\end{theorem}

We give a detailed proof of Theorem~\ref{thm:two} in Appendix~\ref{sec:verification}. We should mention that another potential means of verifying the classical intractability of our sampling protocol would have been to directly measure the $\bigo(n^4)$ local stabilizers of our resource state $\prepstate$, analogous to the technique used in \cite{gao2017quantum, bermejovega2017architectures}. The idea behind this verification scheme is that, if we guarantee our MQC resource state to be the ideal $\prepstate$, then performing our prescribed Pauli measurements should always generate the ideal sampling states $\sampf$. Unfortunately, this resource state verification scheme doesn't detect errors occurring during preparation measurements, so that even when given an ideal MQC resource state, measurement imperfections during state preparation will still lead to logical errors which harm our output sampling state $\rho_f$. In order for this verification scheme to rigorously guarantee the classical intractability of sampling in our setting, the single-qubit error rates for measurement must be less than $\bigo(n^{-4})$, whereas our verification technique only needs errors rates of $\bigo(n^{-1})$. Since this latter rate is the maximum allowed for any kind of sampling to maintain a constant variational error, this shows our verification scheme to be optimal with regards to its soundness under measurement imperfections. The techniques used to achieve these more favorable allowed error rates fundamentally rely on our use of Conjecture~\ref{conj}, and cannot be directly transferred to other sampling settings such as \cite{gao2017quantum, bermejovega2017architectures}.

\section{Outlook}
\label{sec:outlook}

We have demonstrated the use of MQC to perform classically intractable sampling and verification in a unified manner, with identical resource requirements for each task. This shows that verifying the hardness of a quantum sampling protocol doesn't need to be any harder than the actual sampling, and in certain architectures comes essentially for free. This contrasts sharply with many existing quantum supremacy proposals\cite{aaronson2013computational, boixo2016characterizing, fefferman2017exact, deshpande2017complexity}, for which verifying the non-classical nature of sampling is significantly harder than the sampling itself, likely requiring exponential computational resources to ensure correctness. By using nonadaptive MQC to drive our protocol, we have furthermore allowed both sampling and verification to be carried out in constant time, which minimizes the effect of environmental decoherence, and potentially allows us to avoid the use of quantum error correction.

As an outlook, we expect that a hybrid MQC sampling platform combining the simple physical implementation of \cite{gao2017quantum} or \cite{bermejovega2017architectures} with the convenient theoretical analysis and flexibility available here would represent an extremely appealing framework for implementing classically intractable sampling. In particular, a sampling protocol using nonadaptive MQC with non-Clifford $\sqrt{CZ}$ gadgets embedded in a 2D brickwork-type lattice could potentially demonstrate quantum supremacy in constant time using only $\bigo(n \log(n))$ qubits, and with entirely local interactions. Such a protocol would implement the ``sparse" IQP circuits appearing in \cite{bremner2016achieving}, which require only $\bigo(n \log(n))$ two-body interactions. While this can be implemented in our framework using a 2D lattice of $\bigo(n^2 \log(n))$ qubits which generalizes our $\prepstate$, the possibility of reducing resource requirements further, potentially to $\bigo(n \log(n))$ qubits, would require using local complementation operations on graph states. As these operations can quickly generate long-range entanglement using only local $Y$ basis measurements, we consider such capabilities to represent a unique feature of MQC which are well-suited to reproducing the long-range, low-depth quantum circuits often utilized for quantum sampling.

\section{Acknowledgments}
This work was supported in part by National Science Foundation grants PHY-1314955, PHY-1521016, and PHY-1620651.

\appendix

\section{Comparison to Previous Work}
\label{sec:comparison}

We now discuss the relationship of our work to previous proposals for classically intractable sampling with qubits, the class of boson sampling protocols having a largely different flavor with regards to theoretical underpinnings and experimental implementations. As mentioned before, our work is most closely related to that of \cite{bremner2016average}, as it implements their circuit-based IQP sampling in the context of MQC. We have seen that this translation has several practical advantages, mainly that it allows us to use constant depth quantum circuits generated by local interactions to perform classically intractable sampling in constant time. This translation also reveals the role of MQC byproduct operators in simplifying our protocol, with an associated randomness which ends up having no impact on the classical intractability of sampling. Furthermore, the convenient verification scheme utilized in our protocol can be applied equally well in any classically intractable sampling implementation using the IQP sampling states associated with Conjecture~\ref{conj}, revealing an inherent practical advantage of sampling from this class of states. This advantage more generally applies to any protocol which samples from output distributions defined by so-called hypergraph states\cite{rossi2013quantum, guhne2014entanglement}.

Although our work doesn't make use of the alternate Conjecture~2 of \cite{bremner2016average}, concerning the average-case hardness of estimating fully-connected Ising partition functions, our techniques can be easily generalized to define a similar MQC sampling protocol which relies upon Conjecture~2. In this alternate protocol, our $CCZ$ gadget would be replaced by gadgets for the non-Clifford $\sqrt{CZ}$ and $T$ gates, and our byproduct group would contain not only $CZ$, but also $\sqrt{Z}$ gates. In terms of the Clifford hierarchy of unitary operations \cite{gottesman1999demonstrating}, the pattern which emerges here is that using gadgets which implement operations at the third level of the Clifford hierarchy leads to a random byproduct group formed from Clifford gates at the second level of the Clifford hierarchy. Just as with our protocol, this would eliminate the need to apply any Clifford gates ``by hand", reducing the physical resources needed for sampling.

Our work also has many similarities to the MQC sampling protocol of \cite{gao2017quantum}, which similarly runs in constant time using a fixed ``brickwork" resource state preparable by a constant depth quantum circuit, and also allows for verification. In our protocol, the average-case hardness of sampling relies on Conjecture~\ref{conj}, while the average-case hardness in \cite{gao2017quantum} relies upon a conjecture regarding the estimation of output probabilities of random quantum circuits, argued in \cite{lund2017sampling} to be a stronger assumption. On the other hand, this latter conjecture is very similar to that used in \cite{boixo2016characterizing, aaronson2016complexity, bermejovega2017architectures}.

While \cite{gao2017quantum} also achieves verification of the hardness of their sampling distribution, their method requires verifying the entire initial MQC resource state. By contrast, our use of Conjecture~\ref{conj} lets us perform verification in exactly the same manner as sampling, where the only difference is a change in the single-qubit Pauli bases used to perform the final $n$ measurements. This unique duality between sampling and verification arises from the simple byproduct group appearing in our protocol, which is necessary for our preparation measurements to always implement IQP circuits. In contrast, the output states of general random unitary circuits studied in \cite{aaronson2013computational, boixo2016characterizing} are likely too complicated to allow the associated sampling and verification protocols, or MQC counterparts such as \cite{gao2017quantum, bermejovega2017architectures}, to achieve the duality we observe here.

\section{Randomness of MQC Byproduct Polynomials}
\label{sec:preparation}

Here we study the preparation stage of our MQC protocol, and show that the polynomials $f=\fa+\fbc$ associated with our random output states $\rho_f$ contains uniformly random quadratic and linear coefficients, so that every $b_{ij}$ and $c_i$ is an independent binary random variable with equal $\onehalf$ probability. We show this by first characterizing the distribution of preparation outcomes $P_{\fa}(\tvec)$, where $\tvec=(t_1,t_2,\ldots,t_m)$, then using this to characterize the distribution $P_{\fa}(\fbc)$ of ``byproduct polynomials" arising in our protocol. We show that $P_{\fa}(\fbc)$ is uniformly random, a fact which holds true in the presence of arbitrary noise with spatial correlations of a bounded distance. This result will be used in our proofs of sampling and verification in Appendices~\ref{sec:sampling} and \ref{sec:verification}.

We calculate $P_{\fa}(\tvec)$ using the Born rule, which in our ideal setting says that given $\fa$-dependent preparation measurements on $\prepstate$, the probability of obtaining an outcome $\ket{\tvec_{\fa}}$ (where $\fa$ denotes the appropriate single-qubit eigenbases) is
\begin{equation}
    \label{eq:t_prob}
    P_{\fa}(\tvec)=|\braket{\tvec_{\fa}}{\preplabel}|^2.
\end{equation}
The expression $\braket{\tvec_{\fa}}{\preplabel}$ here denotes not a scalar, but a partial inner product on $\prepstate$, consisting of an $n$-qubit state which isn't measured until the sampling and verification stage of our protocol. Consequently, \eqnref{eq:t_prob} says that $P_{\fa}(\tvec)$ is equal to the squared norm of this state $\braket{\tvec_{\fa}}{\preplabel}$. Although we would expect this output state to be the sampling state $\sampf$, a careful calculation of the inner products arising in our protocol reveals an additional $\sqrtonehalf$ scalar factor per preparation measurement, as remarked in Figure~\ref{fig:gadgets}. This shows that $\braket{\tvec_{\fa}}{\preplabel}=(\sqrtonehalf)^m\sampf$, where $f=\fa+\fb(\tvec)+\fc(\tvec)$, which then proves the preparation measurement outcomes to be distributed as $P_{\fa}(\tvec)=2^{-m}$. We note that this independence of measurement outcomes is a generic feature of MQC state preparation protocols, as the implementation of norm-preserving unitary operations in every preparation measurement will necessarily force \eqnref{eq:t_prob} to take a constant value for all $\tvec$, corresponding to every preparation outcome $t_i$ being uncorrelated and uniformly random.

We now use the uniform randomness of preparation measurement outcomes $\tvec$ to prove the uniform randomness of byproduct polynomials $\fbc$, which depend on $\tvec$ as $\fb(\tvec)+\fc(\tvec)$. These global byproducts arise from the local byproduct operators associated with random outcomes $t_i$ in each of the MQC gadgets shown in Figure~\ref{fig:gadgets}, which are then commuted through our computation to contribute linear and quadratic terms to $\fb(\tvec)+\fc(\tvec)$. Each quadratic and linear coefficient in $\fbc$ can thus be expressed as a sum (mod 2) of many different measurement outcomes $t_i$, and it is clear that the complete randomness of each measurement outcome will lead every byproduct coefficient in $\fbc$ which contains even a single random $t_i$ to be itself completely random. It is clear that every quadratic coefficient contains contributions from at least one random $t_i$, with the one exception of $b_{1n}$. Because our $CCZ$ gadgets only apply $CCZ$ byproduct operators between nearest neighbor logical qubits, and since qubits 1 and $n$ are never adjacent to each other in the circuit diagram of Figure~\ref{fig:protocol}, it remains possible that $b_{1n}$ will always be 0. A simple fix for this is to simply vary the ordering among each triple of qubits entering a non-Clifford gadget using $SWAP$ gadgets, so that all qubits are adjacent to all other qubits equally often. In this case, every quadratic coefficient $b_{ij}(\tvec)$ in $\fb(\tvec)+\fc(\tvec)$ will receive $\bigo(n)$ random contributions from outcomes $t_i$ arising in $CCZ$ gadgets, and every linear coefficient $c_i(\tvec)$ will receive $\bigo(n^3)$ contributions from outcomes arising in 1D cluster wires and $SWAP$ gadgets. This clearly proves that the distribution of byproduct operators will be uniformly random as $P_{\fa}(\fbc)=2^{-(\nb+n)}$, where $\nb={n\choose2}$.

The above analysis which counts the number of measurement outcomes contributing to each coefficient of $\fbc$ is unnecessary in an idealized setting, but is useful in the presence of realistic noise and experimental imperfections. We can generally characterize this behavior as a trace preserving quantum operation $\mathcal{E}$ which maps our MQC resource state to some imperfect $\mathcal{E}(\ketbrad{\preplabel})$. Our measurement statistics $P_{\fa}(\tvec)$ in this setting are again set by the Born rule, but now as 
\begin{align}
    \label{eq:t_prob_mixed}
    P_{\fa}(\tvec) &= \Tr\left[\mathcal{E}(\ketbrad{\preplabel}) \ketbrad{\tvec_{\fa}}\right] \\
    &= \Tr\left[\ketbrad{\preplabel} \mathcal{E}^\dagger(\ketbrad{\tvec_{\fa}})\right],
\end{align}
where $\mathcal{E}^\dagger$ represents the quantum operation which is adjoint to $\mathcal{E}$. While $\mathcal{E}^\dagger$ may modify our measurement projectors $\ketbrad{\tvec_{\fa}}$ so as to displace or correlate the probabilities of local outcomes $t_i$, we noted above that the coefficients of byproduct polynomials are determined by at least $\bigo(n)$ different such measurement outcomes, any one of which is capable of completely randomizing the probability of that coefficient. Consequently, in order for noise to alter the distribution of byproduct operators, the operator $\mathcal{E}^\dagger$ must induce correlations between at least $\bigo(n)$ different measurement outcomes in our system. In the presence of any noise with a finite correlation length, this is clearly impossible, which proves the uniform randomness of byproduct operators to be a robust property of our MQC protocol.

\section{Hardness of Approximate Sampling}
\label{sec:sampling}

Here we give a detailed proof of the classical intractability of our MQC sampling protocol under constant variational noise in the output sampling distributions $Q_f$. We first discuss the general idea behind average-case classically intractable sampling protocols, so as to make clear what precisely needs to be demonstrated in our proof. We then describe the use of classical post-processing on our measurement records to implement ``coarse-graining" in the description of our protocol. This coarse-graining lets us simplify the analysis of failure probabilities required in our proof, and eventually lets us prove Theorem~\ref{thm:one}, with its associated variational error threshold of $\eta_0=\oneeightsix$. We note a certain duality between the proof given here and the proof of Theorem~\ref{thm:two} given in Appendix~\ref{sec:verification}, with the former using a guaranteed bound on $\eta_0$ as a starting point and the latter deriving such a bound on $\eta_0$ as an end result.

Any proof of classical intractability of quantum sampling requires adopting somewhat of a dual viewpoint. On the one hand, we recognize that our sampling procedure is an intrinsically quantum task, but at the same time assume that the sampling distributions arising from this quantum process can be exactly replicated using a probabilistic classical algorithm. This assumption, analogous to the assumption of a hidden variable model describing our quantum process, is made in order to derive a (widely conjectured) contradiction, the collapse of the polynomial hierarchy of complexity theory. Even though the probabilities of individual sampling outcomes $Q_f(\svec)$ are exponentially small and would require exponential time to estimate empirically, if they arise from a classical sampling process, then the technique of Stockmeyer approximate counting can be used to estimate these probabilities up to multiplicative error. In particular, Stockmeyer counting can be used to output an estimate $\qest(\svec)$ which is related to our probability of interest by $|\qest(\svec)-Q_f(\svec)|\leq \frac{Q_f(\svec)}{\poly(n)}$, for any desired polynomial $\poly(n)$. The use of an average-case complexity conjecture, like Conjecture~\ref{conj} in our paper, is then required to connect the ability to estimate such probabilities in the presence of noise to the ability to solve \#P-hard problems, from which a collapse of the polynomial hierarchy follows.

Stockmeyer counting is an unphysical process which cannot be carried out efficiently using classical or quantum devices, but can be implemented with a hypothetical ``alternating Turing machine" capable of efficiently solving problems in the third level of the polynomial hierarchy \cite{footnote2}. Furthermore, Stockmeyer counting involves manipulations on a register of binary random variables underlying our random outcomes, and consequently can only estimate probabilities arising as outcomes of classical randomized algorithms. Nonetheless, if we assume the existence of an efficient classical algorithm for exactly sampling from the distribution $D_f(\svec)=\ngapsq(f+\svec)$, Stockmeyer sampling would then permit a device existing in the third level of the polynomial hierarchy to estimate any $\ngapsq(f)$ up to multiplicative error, and thus solve any problem in \#P. Because solving arbitrary problems in \#P is known by Toda's theorem \cite{toda1991pp} to allow one to efficiently solve all problems in the hierarchy, assuming the existence of this efficient classical algorithm for sampling from distributions $D_f$ would necessarily collapse the polynomial hierarchy to its third level, a contradiction. Hence, this proves the task of sampling from arbitrary $D_f$ to be classically intractable.

A necessary ingredient in any \textit{average-case} classically intractable sampling result is a mathematical problem whose estimation remains \#P-hard even when our estimates have some finite probability of failing to be multiplicatively close to their actual value. In our setting, this problem is furnished by Conjecture~\ref{conj}, which says that estimating $\ngapsq(f)$ up to $\onequarter$ multiplicative error is \#P-hard, even when a fraction $\epsilon\leq \epsilon_0=\frac{23}{24}$ of our estimates fail to lie within this $\onequarter$ multiplicative bound. Evidence in support of Conjecture~\ref{conj} is given in \cite{bremner2016average}. This failure probability $\epsilon_0$ ends up determining the allowed deviation of our quantum sampling distributions $Q_f$ from their ideal $D_f$. If this deviation is sufficiently small, as measured by the variational distance between $Q_f$ and $D_f$, the assumed computational hardness of estimating $\ngapsq(f)$ then guarantees that our quantum sampling task will be classically intractable. Consequently, our main goal in this proof is to analyze the deviations in our distributions $Q_f(\svec)=\Tr(\rho_f\ketbrad{\svec_X})$ arising from deviations in our experimental states $\rho_f$ from their ideal $\ketbrad{\psi_f}$, and to find sufficient conditions to guarantee that the failure probability in estimating $\ngapsq(f)$ using Stockmeyer sampling on $Q_f$ is below our threshold $\epsilon_0$.

We now introduce the idea of coarse-grained sampling distributions, which indeed we have already implicitly made use of in the description of our sampling protocol. In Section~\ref{sec:protocol}, we described different preparation outcomes $\tvec=(t_1,t_2,\ldots,t_m)$ as giving rise to different ideal sampling states $\ket{\psi_{f(\tvec)}}$ via the correspondence $f(\tvec)=\fa+\fb(\tvec)+\fc(\tvec)$. This means that whenever different preparation outcomes $\tvec\neq\tvec'$ generate the same byproduct polynomials $\fb(\tvec)+\fc(\tvec)=\fb(\tvec')+\fc(\tvec')$, the resultant sampling states will be identical. In reality though, it is entirely possible that these preparation outcomes will generate different sampling states $\rho_{\fa,\tvec}\neq\rho_{\fa,\tvec'}$, leading our description of a single sampling state $\rho_{f(\tvec)}$ to represent a coarse-graining over equivalent preparation outcomes $\tvec$. In particular, if $P_{\fa}(\tvec)$ denotes the probability of obtaining a preparation outcome $\tvec$ arising from our $\fa$-dependent Pauli measurements on $\prepstate$, then we find $\rho_f$ to be given by
\begin{equation}
    \rho_f = \frac{1}{P_{\fa}(\fb,\fc)}\sum_{\{\tvec|\fb(\tvec)+\fc(\tvec)=f+\fa\}} P_{\fa}(\tvec)\rho_{\fa,\tvec}\,\,.
\end{equation}
$P_{\fa}(\fb,\fc)$ represents a normalization factor which gives the total probability on input $\fa$ of obtaining any outcome $\tvec$ associated with the byproduct polynomial $\fbc=f+\fa$. While the above coarse-graining might appear trivial, we will now show how this can be used to effectively mix the inequivalent states $\rho_f$ and $\rho_{f'}$ when $f$ and $f'$ differ only in their linear coefficients.

If we describe our overall sampling process at this stage as first preparing a random state $\rho_f$ with $f=\fa+\fb+\fc$, which is then sampled to obtain an $X$ basis outcome of $s$, then we would record this in an experiment as yielding the outcome $(\fb,\fc,\svec)\in \Omega_{\fa}$ in some outcome space $\Omega_{\fa}$. From the layout of our sampling protocol, the probability of this outcome is clearly $P_{\fa}(\fb,\fc,\svec)=P_{\fa}(\fb,\fc)Q_{f}(\svec)$. Because of the degeneracy $D_{f+\svec}(\svec)=\ngapsq(f)$ for all outcomes $\svec$, we say that any such outcome samples from the polynomial $f$. These exponentially many outcomes are precisely the ones which can be used to obtain an estimate of $\ngapsq(f)$ via Stockmeyer counting, and we will choose our coarse-graining to eliminate this degeneracy, so that each $\ngapsq(f)$ is determined by a unique sampling outcome from a unique output sampling state. We note that this coarse-graining was used implicitly in \cite{bremner2016average}, although interpreted there as an ``obfuscation" of output probabilities.

In Appendix~\ref{sec:preparation} we showed that the distribution of byproduct polynomials is uniformly random as $P_{\fa}(\fb,\fc)=2^{-(\nb+n)}$, where $\nb={n\choose2}$. Given this robust characterization of $P_{\fa}(\fb,\fc)$, we will use $\qcoarse(\fc)$ to indicate the conditional probability of obtaining any outcome which samples from $f=\fab+\fc$, given that the quadratic portion of our byproduct polynomial is $\fb$. This leads $\qcoarse(\fc)$ to be
\begin{align}
    \label{eq:rho_ab}
    \qcoarse(\fc) &= 2^{\nb}\sum_s P_{\fa}(\fb,\fc+\svec) Q_{f+\svec}(\svec) \\
    &= \sum_{\svec} 2^{-n} \Tr\left(\rho_{f+\svec} \ketbrad{\svec_X}\right) \\
    &= \Tr\left(2^{-n} \sum_{\svec}\rho_{f+\svec} \ketbrad{\svec_X}\right) \\
    &= \Tr\left(\rhocoarse \ketbrad{{\fc}_{X}}\right). \label{eq:prob_1}
\end{align}
We use $\ket{{\fc}_{X}}$ to indicate the $X$ basis outcome string corresponding to the linear terms of $f$. In the above, we have also defined $\rhocoarse$ to be the state
\begin{equation}
    \label{eq:rho_1}
    \rhocoarse = 2^{-n} \sum_{\svec} Z^{\svec}\left(\rho_{\fab+\svec}\right)Z^{\svec},
\end{equation}
where $Z^{\svec}=\bigotimes_{i=1}^n (Z_i)^{s_i}$ indicates a product of $Z$ operators. In the ideal setting where each $\rho_f=\ketbrad{\psi_f}$, the result of applying $Z^{\fc}$ to $\rho_{f}$ is to simply remove the linear components of $f$, leaving the state $\ketbrad{\psi_{\fab}}$. In this idealized setting, the result of averaging over all $\rho_{f}$ and applying the correction $Z^{\fc}$ in each case is to leave the state $\rhocoarse=\ketbrad{\psi_{\fab}}$, which contains only cubic and quadratic terms. While we can't literally implement these unitary corrections $Z^{\fc}$ within the setting of MQC, we can simulate their action through classical postprocessing on our measurement outcomes. In particular, whenever we obtain an outcome of $(\fb,\fc,\svec)\in \Omega_{\fa}$ in our sampling experiment, we instead record this as a coarse-grained outcome $(\fb,\fc+\svec)\in \tilde{\Omega}_{\fa}$ lying in a simpler outcome space $\tilde{\Omega}_{\fa}$. This is equivalent to recording only the polynomial $f$ sampled by our experiment, and forgetting the relative contributions to $f$ from MQC byproduct operators and from sampling outcomes $\svec$. The equivalence of this coarse-graining in our measurement records with the action of active unitary corrections arises from the equality $\ketbrad{\svec_X} = Z^{\fc+\svec}\ketbrad{{\fc}_{X}}Z^{\fc+\svec}$ used to derive \eqnref{eq:prob_1}.

Given this coarse-grained description of our experiment, we would like to bound the failure probability $\epsilon$ of obtaining an estimate $\ngapsqest(f)$ which differs from the true $\ngapsq(f)$ by more than a multiplicative factor of $\onequarter$. By requiring this probability to be less than the $\epsilon_0=\frac{23}{24}$ appearing in Conjecture~\ref{conj}, we will arrive at concrete conditions on our coarse-grained output states $\rhocoarse$ in order for our MQC protocol to implement classically intractable sampling. While the Stockmeyer counting used to obtain $\ngapsqest(f)$ from our sampling probabilities $\qcoarse(\fc)$ technically introduces its own multiplicative error in this estimate, because this error can be reduced in our (hypothetical) Stockmeyer counting algorithm to any inverse polynomial $|\ngapsqest(f)-\qcoarse(\fc)|<\frac{\qcoarse(\fc)}{\poly(n)}$ while still retaining a polynomial runtime, we will ignore this error in the following and simply set $\ngapsqest(f)=\qcoarse(\fc)$.

We first use Markov's inequality to bound the probability of our estimate $\ngapsqest(f)$ failing to lie within an arbitrary constant distance of $\ngapsq(f)$, $\Prob_{f}\left(|\ngapsqest(f)-\ngapsq(f)|>2^{-n} \delta\right)$, over arbitrary polynomials $f=\fab+\fc$. We will later convert this into a failure probability for obtaining an estimate of $\ngapsq(f)$ outside of our allowed $\onequarter$ multiplicative error. Since the approximate and exact values of $\ngapsq(f)$ can both be interpreted as probabilities in different distributions, $\ngapsqest(f)=\qcoarse(\fc)$ and $\ngapsq(f)=D_{\fab}(\fc)$, we find that the distance between these values, averaged over $\fc$ with fixed $\fab$, is proportional to the variational distance between these distributions as
\begin{align}
    \average{|\ngapsqest(f)-&\ngapsq(f)|}_{\fc} \nonumber\\
    &= 2^{-n}\sum_{\fc}|\qcoarse(\fc)-D_{\fab}(\fc)| \\
    &= 2^{-n}\onenormc{\qcoarse-D_{\fab}}
\end{align}
Defining $\eta_{\fab}=\onenormc{\qcoarse-D_{\fab}}$ to be the variational distance between these distributions, Markov's inequality then tells us that for any $\delta>0$ and for $f=\fab+\fc$ with a fixed $\fab$,
\begin{align}
    \label{eq:markov}
    \Prob_{\fc}\left(|\ngapsqest(f)-\ngapsq(f)|>2^{-n}\delta\right) < \frac{\eta_{\fab}}{\delta},
\end{align}

Having this bound in hand, we now give an anticoncentration bound on the probability that $\onequarter\ngapsq(f)<2^{-n}\delta$, which lets us convert the above bound into a statement about the failure probability $\epsilon$. We utilize a particular form of Cantelli's inequality stating that for any non-negative random variable $X$ and constant $\delta'$ in $0\leq\delta'\leq1$,
\begin{equation}
    \Prob(X\leq\delta'\average{X}) \leq \frac{\average{X^2}-\average{X}^2}{\average{X^2}-\delta'(2-\delta')\average{X}^2}.
\end{equation}
This agrees with the more well-known Paley-Zygmund inequality at $\delta'=0,1$, but otherwise gives a more stringent upper bound. Setting $X=\ngapsq(f)$, $\delta'=4\delta$, and using the result $\average{\ngap^4(\fa+\fbc)}_{\fb,\fc}\leq3\cdot2^{-2n}$ from \cite{bremner2016average}, this lets us restrict the probability of $\onequarter\ngapsq(f)$ being less than $2^{-n}\delta$ as
\begin{align}
    \label{eq:pz}
    \Prob_{\fb,\fc}\left(\onequarter\ngapsq(\fa+\fbc)\leq2^{-n}\delta\right) \leq \frac{2}{2+(1-4\delta)^2}.
\end{align}
We now define $\eta=\average{\eta_{\fab}}_{\fa,\fb}$ to be the average variational distance between distributions $\qcoarse$ and $D_{\fab}$, averaged over all $\fab$. Combining \eqnref{eq:markov} with the average of \eqnref{eq:pz} over all $\fa$, this results in a bound on the multiplicative failure probability of
\begin{multline}
    \Prob_{f}\left(|\ngapsqest(f)-\ngapsq(f)|>\onequarter\ngapsq(f)\right) \\
    < \frac{\eta}{\delta} + \frac{2}{2+(1-4\delta)^2},
\end{multline}
which holds for every $0\leq\delta\leq\onequarter$.

We now require the failure probability to be at most $\frac{23}{24}$, in line with Conjecture~\ref{conj}, and numerically optimize over $\delta$ to find the largest allowed value of $\eta_0$ for which this can be achieved. This yields a maximum of $\eta_0=0.01169$, which has a rational lower bound of $\eta_0\approx\oneeightsix$. This completes our proof of Theorem~\ref{thm:one}.

\section{Verification of Classical Intractability}
\label{sec:verification}

Here we prove that the verification scheme occurring in the last stage of our MQC protocol does indeed guarantee the classically intractable of our sampling process. We first show that the local $X$ and $Z$ measurements made on our sampling states $\rho_f$ during verification correspond to exact measurements of the nonlocal stabilizers $\hami$, via the parity functions $\pami(\vvec)$. This allows us to estimate the average $\average{\hami}_{i,f}$ with respect to random $\rho_f$, which allows us to bound the average variational distance $\average{\onenormc{Q_f-D_f}}_f$ using results from \cite{hangleiter2017direct}. If our empirical estimate of $\average{\hami}_{i,f}$ remains sufficiently low, an application of H{\"o}ffding's inequality lets us show that $\bigo(n^2)$ verification measurements are sufficient to conclude that $\average{\onenormc{Q_f-D_f}}_f \leq \oneeightsix$ with any fixed statistical significance, proving Theorem~\ref{thm:two}.

We first briefly review our verification procedure. After preparation of a random $\rho_f$, we choose with 50\% probability to perform either sampling or verification measurements on $\rho_f$. If verification is chosen, we further choose a random qubit $i$ of $\rho_f$ which is measured in $X$, while all other $n-1$ qubits are measured in $Z$. We denote the measurement outcome string by $\vvec=(v_1,v_2,\ldots,v_n)$, ignoring the fact that $v_i$ is associated with a different measurement basis. We then use our knowledge of the polynomial $f$ associated with $\rho_f$ to compute a parity function of $\vvec$, $\pami(\vvec) = \partial_{i}f(\vvec)+v_i$, where $\partial_i f$ is the polynomial difference $\partial_{i}f(\vvec) = f(v_1,\ldots,v_i+1,\ldots,v_n)-f(v_1,\ldots,v_i,\ldots,v_n)$. It is easy to show that $\partial_{i}f(\vvec)$ is independent of the value of $v_i$.

We show here that the process of measuring $\vvec$ using single-qubit Pauli measurements and then computing $\pami(\vvec)$ is exactly equivalent to measuring the nonlocal stabilizer $\hami$ as $\hami(\vvec)=(-1)^{\pami(\vvec)}$, where $\hami(\vvec)$ indicates the $\hami$ outcome corresponding to $\vvec$. Both processes yield binary random variables as their output, and in order to prove that their probability distributions are identical, we can prove that both measurement schemes are associated with identical Hermitian observables. While measurements of $\hami$ are clearly associated with the Hermitian operator $\hami$ itself, it isn't immediately clear how we should interpret the measurements of $\vvec$ as measuring any particular Hermitian operator. The answer comes by recognizing that our relevant measurement statistics during verification consist only of the binary values $\pami(\vvec)$, and forgets the specific outcomes $\vvec$ which produced them. Translating these $\pami$ outcomes into equivalent $\hami$ outcomes shows the expectation value of $\hami$ on $\rho_f$ to be
\begin{align}
    \label{eq:pami_hami}
    \average{(-1)^{\pami(\vvec)}}_{\vvec} &= \sum_{\vvec\in GF(2)^n} (-1)^{\pami(\vvec)} \Prob(\vvec|\rho_f) \\
    &= \sum_{\vvec} (-1)^{\partial_{i}f(\vvec)+v_i} \Tr\left[\rho_f \left(H_i\ketbrad{\vvec}H_i\right)\right] \\
    &= \Tr\left[\rho_f \left( X_i \sum_{\vvec} (-1)^{\partial_{i}f(\vvec)} \ketbrad{\vvec}\right)\right] \\
    &= \Tr\left(\rho_f \hami\right).
\end{align}
In the last equality, we have used the definition of $\hami$ in \eqnref{eq:hami}, while in the second to last equality we used $X_i=\sum_{v_i} (-1)^{v_i} H_i\ketbrad{v_i}H_i$. This reveals that the expectation value of $(-1)^{\pami(\vvec)}$ is equal to that of $\hami$ on $\rho_f$, and since we made no assumptions about $\rho_f$, this shows that our verification scheme is exactly equivalent to measuring $\hami$ \cite{footnote3}.

As a concrete example, suppose we are working with the 3-qubit sampling state $\ket{\psi_{x_1 x_2 x_3}}=CCZ_{123}\ket{+}^{\otimes 3}$ and wish to measure the stabilizer $h_{x_1 x_2 x_3}^{(1)}=X_1 CZ_{23}$. In this case, we would perform our verification by measuring $X$ on qubit 1, $Z$ on qubits 2 and 3, and then computing the polynomial $\pami(\vvec)=v_1 + v_2 v_3$. This process, which can be thought of as obtaining classical values and plugging them in to the stabilizer itself, would indicate a success when $v_1=1$ and $v_2=v_3=1$, or when $v_1=0$ and at least one of $v_2=0$ or $v_3=0$ holds true.

Given the ability to measure arbitrary $\hami$ using single-qubit $X$ and $Z$ measurements, we now note that the average $\average{\hami}_{i} = \frac{1}{n}\sum_i \average{\hami}$ over randomly chosen sites $i$ is equal to 1 on a given $\rho_f$ only when $\rho_f=\ketbrad{\psi_f}$ is the ideal sampling state. More generally, the techniques of \cite{hangleiter2017direct} show that this average can be used to bound the closeness of $\rho_f$ to $\ketbrad{\psi_f}$, as measured by the fidelity $F_f=\sqrt{\matrixel{\psi_f}{\rho_f}{\psi_f}}$. For our purposes, it will be more convenient to work with the square of this quantity, $F_f^2$. When $\average{\hami}_{i}\geq 1-\frac{2}{n}$, $\rho_f$ cannot be orthogonal to $\sampf$, and must have a fidelity squared of at least $F_f^2 \geq 1-\frac{n}{2}(1-\average{\hami}_{i})$. If we average both sides of this equality over polynomials $f=\fa+\fbc$ with random $\fbc$, then we find that the average fidelity squared $\average{F_f^2}_f$ of output states $\rho_f$ relative to their intended $\sampf$ is bounded by the average $\average{\hami}_{i,f}$ as
\begin{equation}
    \label{eq:fidelity_bound}
    \average{F_f^2}_f \geq 1-\frac{n}{2}(1-\average{\hami}_{i,f})
\end{equation}

With \eqnref{eq:fidelity_bound} in hand, we can now bound the average variational distance $\average{\onenormc{Q_f-D_f}}_f$ between the sampling distributions arising from $\rho_f$ and $\sampf$. We utilize the fact that the quantum 1-norm distance $\onenormq{\rho_f-\ketbrad{\psi_f}}\geq \onenormc{Q_f-D_f}$ gives an upper bound on the variational distance of any output sampling distributions, where $\onenormq{\rho_f-\ketbrad{\psi_f}} = \Tr\left(\big|\rho_f-\ketbrad{\psi_f}\big|\right)$ with $|A|$ the operator absolute value. We also use a well-known bound on the 1-norm distance, $\onenormq{\rho_f-\ketbrad{\psi_f}}\leq \sqrt{1-F_f^2}$, which together yield
\begin{align}
    \average{\big|Q_f-D_f\big|_1}_f &\leq \average{\onenormq{\rho_f-\ketbrad{\psi_f}}}_f \\
    &\leq \left\langle\sqrt{1-F_f^2}\,\right\rangle_f \label{eq:bound_one}\\
    &\leq \sqrt{1-\average{F_f^2}_f} \label{eq:bound_two}\\
    &\leq \sqrt{\frac{n}{2}(1-\average{\hami}_{i,f})}. \label{eq:bound_three}
\end{align}
In the above, we used the two bounds mentioned, as well as Jensen's inequality for the concave function $\sqrt{1-X}$ in \eqnref{eq:bound_two}. Using the relationship between the average of stabilizers and parity functions, $\average{\hami}_{i,f}=\average{(-1)^{\pami(\vvec)}}_{\vvec,i,f}=1-2\average{\pami(\vvec)}_{\vvec,i,f}$, this finally lets us show that in order to verify that $\average{\big|Q_f-D_f\big|_1}_f\leq\eta_0$, it is sufficient for our parity function average to be below
\begin{equation}
    \label{eq:empirical_bound}
    \average{\pami(\vvec)}_{\vvec,i,f} \leq \frac{\eta_0^2}{n}.
\end{equation}
This gives the bound appearing in Theorem~\ref{thm:two}.

Although any empirical estimate of $\average{\hami}_{i,f}$ obtained from finitely many measurements of $\pami(\vvec)$ isn't guaranteed to accurately reflect its true value, we can bound the closeness of this estimate with high probability using the uniformly random distribution of byproduct operators proved in Appendix~\ref{sec:preparation}. In particular, this tells us that for any fixed $\fa$, the average $\average{\hami}_{i,\fb,\fc}$ over output random byproducts is unbiased towards any fixed $\rho_f$, and thus is an accurate indicator of the uniform closeness of sampling states. This lets us treat $\average{\hami}_{i,f}$ as a simple binary random variable, and use H{\"o}ffding's inequality to bound the probability of this estimate deviating too far from the true value of $\average{\hami}_{i,f}$. 

H{\"o}ffding's inequality says that if we obtain an estimate $\tilde{X}$ of a binary random variable $X$ using $N$ independent samples, the probability of the true average $\average{X}$ lying above $\tilde{X}$ by more than $\zeta$ is
\begin{equation}
    \label{eq:hoeffding}
    \Prob(\average{X}\geq\tilde{X}+\zeta) \leq \exp(-2\zeta^2 N)
\end{equation}
In our case, we choose $X$ to be our random parity function, and $\zeta$ to be the difference between our specified tolerance $\frac{(1/86)^2}{n}$, and the more numerically precise tolerance for classically intractable sampling derived in Appendix~\ref{sec:sampling}, $\frac{(0.01169)^2}{n}$. Setting $N=\mu n^2$, this gives a failure probability of
\begin{equation}
    p_F\leq\exp(-(2.9138\times10^{-6}) \mu^2) = \exp(-\bigo(\mu^2)).
\end{equation}
Converting this into a success probability $p=1-p_F$ then completes our proof of Theorem~\ref{thm:two}.

A final remark is given to our means of measuring the highly nonlocal, non-Pauli stabilizers $\hami$ through single-qubit Pauli measurements. This technique can actually be generalized to measure the stabilizers of any sampling state formed by starting with $\ket{+}^{\otimes n}$ and applying an IQP circuit composed of $\sqrt{Z}$, $Z$, $CZ$, $CCZ$, and any higher multiply-controlled $Z$ gates. Since these states include all hypergraph states as special instances, our means of measuring non-Pauli stabilizers can be utilized for the goal of measuring hypergraph stabilizers in \cite{morimae2017verified}, as the latter requires the non-Pauli portion of measured stabilizers to have support on a constant number of qubits. Generalizing yet further, we see that the necessary and sufficient condition for a local measurement scheme to exactly replicate measurements of a nonlocal operator $M$ in this manner is that $M$ can be diagonalized in a basis which is a tensor product of single-qubit eigenbases. While this allows us to measure many different multi-qubit operators using only single-qubit measurements, a simple counterexample is given by the Hermitian operator $SWAP$, which cannot be measured in this manner owing to its unique $-1$ eigenstate being the entangled $\sqrtonehalf(\ket{01}-\ket{10})$.

\begin{thebibliography}{99}
\bibitem{footnote1}
    The proof actually allows for the existence of some multiplicative error, in the form of distributions $\qest$ which satisfy $|\qest(\svec)-Q_{f}(\svec)| < \frac{Q_{f}(\svec)}{\poly(n)}$ for all outcomes $\svec$, with $\poly(n)$ a fixed polynomial. While sampling from such a distribution $\qest$ is still classically intractable, this is unsatisfactory from a practical standpoint. For example, if any outcome $\svec_0$ satisfies $Q_{f}(\svec_0)=0$, then we must have the probability $\qest(\svec_0)$ be exactly 0 as well. This is clearly impossible to verify for any experimental distribution $\qest$, leading exact classically intractable sampling results to have a more strained relationship with experimental realities than their average-case counterparts.
\bibitem{footnote2}
    For comparison, the familiar complexity classes P and NP are respectively contained in the zero and first levels of the polynomial hierarchy (PH). While the randomized complexity class BPP has only been proven to lie in the second level of the PH, a proof of the widely conjectured P=BPP would place it in the zero (lowest) level as well. As a corollary, proving P=BPP would allow Stockmeyer counting to be implemented in the second level of the PH, causing the hypothetical collapse invoked in classically intractable sampling results to occur at the second level of the PH, rather than the third.
\bibitem{footnote3}
    On the other hand, the effect of the measurements used in our verification scheme on the measured state is different from that of direct measurements of $\hami$. For example, performing a genuine quantum nondemolition measurement of $\hami$ on the sampling state $\sampf$ would leave it unchanged, whereas our scheme always collapses it to a tensor product of single-qubit $X$ and $Z$ eigenstates. Since we only care about measurement statistics and not the post-measurement state, this has no impact on our protocol.
\end{thebibliography}
\end{document}